# Adaptive Beamwidth Configuration for Millimeter Wave V2X Scheduling

Baldomero Coll-Perales[1], Javier Gozalvez[1], Esteban Egea-Lopez[2]
[1]UWICORE laboratory, Universidad Miguel Hernandez de Elche, Elche (Alicante), Spain.
[2]GIRTEL group, Universidad Politécnica de Cartagena, Cartagena (Murcia), Spain
[1]{bcoll, j.gozalvez}@umh.es; [2]esteban.egea@upct.es

*Abstract*—Millimeter wave (mmWave) technologies will support the high bandwidth and data rate requirements of V2X services demanded by connected and automated vehicles (CAVs). MmWave V2X technologies will leverage directional antennas that challenge the management of the communications in dynamic scenarios including the identification of available links, beams alignment, and scheduling. Previous studies have shown that these challenges can be reduced when mmWave communications are supported by side information like the one transmitted in sub-6GHz V2X technologies. In this context, this paper proposes a beamwidth-aware mmWave scheduling scheme for V2V communications supported by sub-6GHz V2X technologies. The proposal enables mmWave transmitters to schedule a mmWave transmission to several neighboring vehicles at the same time by adapting the beamwidth configuration. In addition, the proposal derives the minimum beamwidth that mmWave transmitters should use to contact their neighboring vehicles in a limited number of scheduling intervals. The obtained results demonstrate that the proposal helps increasing the amount of mmWave data that can be transmitted to neighboring vehicles.

## I. Introduction

To date, extensive research has been done at the sub-6GHz band based on the IEEE 802.11p and LTE-V2X (a.k.a. Cellular-V2X and LTE-V) standards. IEEE 802.11p- and LTE-based V2X standards utilize the sub-6GHz band to exchange broadcast messages at low data rate that are critical to support active safety services. Sub-6GHz V2X technologies are constrained by their limited bandwidth and are challenged to satisfy the increasing needs of the automotive industry, including the demanding use cases for connected and automated vehicles (CAVs). This has raised the interest on exploring the large bandwidth available at the millimeter wave (mmWave) band (i.e., from 30 GHz to 300 GHz). In addition, automotive radars already use mmWave spectrum, which motivates mmWave automotive joint communication-radar systems [1].

mmWave relies on large antenna arrays and high-directional beams to compensate its severe blockage and pathloss. Directional mmWave transmissions challenge the design of MAC (Medium Access Control) protocols, including the identification of links, beam alignment, and scheduling of mmWave transmissions. The dynamic and fast-changing topologies of vehicular networks increase the mmWave MAC challenges. First evaluations of mmWave MAC for V2V (based on IEEE 802.11ad) have shown high inefficiencies in terms of overhead and latency [2]. A. Loch *et al*. [3] propose to address the beam alignment challenge in V2I scenarios by fixing the mmWave antennas' beams at both the infrastructure and the vehicles. An alternative approach is presented by J. Choi *et al*. in [4] that proposes to exploit side (or out-of-band) information to facilitate the beam alignment in mmWave V2I communications. The authors propose using information received from sub-6GHz V2X communications to facilitate mmWave V2X communications. In particular, the authors presented in [5] and [6] a MAC mechanism for mmWave V2V communications that uses sub-6GHz V2X technologies (based on IEEE 802.11p and LTE, respectively) to support the mmWave beam alignment, identify mmWave links and potential neighbors, and schedule mmWave data transmissions. Current mmWave MAC schemes for V2X communications are designed to schedule transmissions with neighbors sequentially; neighbors are contacted when previous transmissions end. The authors discussed in [5] that beamwidth-aware scheduling solutions could exploit the flexibility in the selection of the beamwidth to schedule mmWave transmissions to several receivers at the same time. [7] builds from this idea and explores the latency benefits of increasing the beamwidth to schedule multiple neighbors simultaneously and reducing the number of transmissions.

Beamwidth-aware scheduling is expected to have a high impact on V2X services where the same data need to be transmitted to all (or a subset of) neighboring vehicles. This is the case for example in Collective Perception Services (CPS) where vehicles exchange sensor data to enlarge their sensing range. However, the design of beamwidth-aware scheduling mechanisms is not exempt of challenges. Solutions need to trade-off beamwidth, antenna gain, interference, and coverage area so that the intended neighboring vehicles can be addressed simultaneously at no reliability cost. The scheduling mechanisms need also to support the simultaneous transmission to multiple vehicles. To the best of the authors' knowledge, no study has yet proposed a scheduling solution that addresses the multiple challenges of the beamwidth-aware mmWave transmissions. In this context, this paper proposes a sub-6GHz assisted beamwidth-aware mmWave scheduling scheme. The proposal relies on the decoupling of the mmWave data and control functions, and offloading mmWave MAC control functions to a sub-6GHz V2X technology [5]. With this approach, the proposal looks for exploiting the longer range, broadcast, and omnidirectional transmissions of sub-6GHz V2X to be used in the control plane, while the directional mmWave transmissions are used in the data plane. This study proposes to exploit the information transmitted in sub-6GHz V2X technologies to identify the location of the neighbors to be addressed, and to derive the minimum beamwidth that mmWave transmissions should use to contact them in a limited number of scheduling intervals. The proposal also integrates the scheduling of the mmWave transmissions to the groups of contacted neighbors using different beams. With this approach, this paper is aimed at studying the potential of beamwidth-aware scheduling solutions for mmWave V2V communications supported by sub-6GHz V2X technologies. The study is also aimed at shedding light into the impact of effects such as varying beamwidths (antenna gains) and interference on the performance of beamwidth-aware mmWave scheduling solutions.

This work has been partially funded by the Spanish Ministry of Science, Innovation, and AEI (PID2020-115576RB-I00, PID2020-112675RB-C41), and FEDER funds (IJC2018-036862-I, TEC2017-88612-R), the Generalitat Valenciana (GV/2021/044) and the Diputación Provincial de Alicante.

## II. BEAMWIDTH-AWARE SCHEDULING FOR MMWAVE V2X

The proposal exploits the information regularly transmitted in broadcast awareness messages (i.e., CAMs –Cooperative Awareness Messages– or BSM –Basic Safety Messages–) of sub-6GHz V2X technologies to identify neighboring vehicles, mmWave links under Line-Of-Sight (LOS) conditions, and directions where the transmitter and receiver should point their beams to perform the beam alignment. mmWave data transmissions are also announced using sub-6GHz V2X broadcast awareness messages. To this aim, mmWave transmitters include in these messages the IDs of the scheduled neighbors, the time instant at which the mmWave data transmission starts, and the duration of the mmWave data transmission. Sub-6GHz V2X broadcast awareness messages are then utilized to announce the scheduling decisions and organize the access to the mmWave channel. The proposed beamwidth-aware scheduling solution allows that mmWave transmitters can schedule transmissions to multiple receivers at the same time. To this aim, a mmWave transmitter can adjust its mmWave antenna's beamwidth. mmWave's antennas are modeled by dividing the horizontal plane into a number of virtual sectors with equal apertures and gain. Beams are then mapped into each of the antenna's sector. Augmenting the beamwidth requires that the number of antenna's sector reduces. This can be achieved by grouping neighboring sectors. For example, a 60-sector antenna would result in beams of 6º width each (i.e., 360º/60). Every two sectors can be grouped to transform the antenna into a 30-sector antenna with beams of 12º width each. Narrower beams allow to concentrate more the antenna's RF energy which produces higher antenna gains. On the other hand, wider beamwidths cover wider areas, and it is then key to increase the number of vehicles that can be reached at the same time. Considering this trade-off, the proposed beamwidth-aware scheduling solution first seeks identifying the thinner beamwidth that a mmWave transmitter should use in order to address its neighboring vehicles in a limited number of scheduling intervals. Then, the proposal schedules the transmission to the different groups of neighbors that are addressed using the different beams at each point in time.

The operation of the proposed beamwidth-aware scheduling scheme is summarized in the pseudo-code shown in Fig. 1. This pseudo-code considers that a mmWave transmitter needs to contact $N$ neighboring vehicles in a scheduling period of $I$ s. The scheduling period is divided into $k$ consecutive scheduling intervals of equal duration $I_i = T_s$ s, $i=1…k$ (**lines 1-2** of Fig. 1). During a period $I$, the mmWave transmitter might have been addressed to receive data from other mmWave transmitters. Then, it has to identify the subset of scheduling intervals where it has not any scheduled transmission ($I_{tx}$) and that it can utilize to schedule transmissions to its neighboring vehicles. The number of scheduling intervals in $I_{tx}$ is represented by the variable $F$ (**line 3** of Fig. 1). In **line 4** of Fig. 1, $S$ represents the set of available beamwidths that the mmWave transmitter can utilize to contact its neighboring vehicles. This study considers that the mmWave transmitter uses the same beamwidth in the available scheduling intervals $I_{tx}$ to contact the neighboring vehicles. $S$ is organized in ascending order (i.e., $s_i < s_j$, if $j > i$, $i=1 …, n$). Then, following this ascending order over $S$, the mmWave transmitter seeks identifying the minimum beamwidth $s^*$ to be used to schedule the transmissions. For the considered $s_i$, the mmWave transmitter first checks the number of beams $B$ that it would need to utilize in order to contact the $N$ neighboring vehicles (**line 6** of Fig. 1). The

```
Algorithm: beamwidth-aware scheduling
1.  N → mmWave neighboring vehicles
2.  I = {I_1, ...I_k} → Set of consecutive scheduling intervals of dur. T_s s
3.  [I_tx, F] = CheckSchedTx() → subset I_tx of size F
4.  S={s_1, ..., s_i, ... s_n} → Set of available beamwidths
5.  For S
6.      B= beams() → Beams needed to contact the N neighbors
7.      if B > F then
8.          Increase beamwidth (i.e., s_i → s_{i+1})
9.          Continue
10.     End if
11.     s* = s_i → Minimum beamwidth identified
12.     [T_tx]=scheduleTx() → transmissions scheduled in I_tx
13.     Break
14. end
```

Fig. 1. Pseudocode of the proposed beam-aware scheduling scheme.

mmWave transmitter exploits for this purpose the location information transmitted by the neighboring vehicles in the sub-6GHz broadcast awareness messages. Note that neighboring vehicles that can be addressed using the same beam are contacted in the same scheduling interval. Then, the mmWave transmitter identifies whether it could schedule a transmission to the $N$ neighboring vehicles if the number of available scheduling intervals $F$ is equal or higher than the number of beams $B$ utilized to contact them (**line 7** of Fig. 1). If this is not the case, the mmWave transmitter realizes that it needs to use a wider beamwidth (**line 8** of Fig. 1). Finally, when the mmWave transmitter identifies the beamwidth $s^*$ (**line 11** of Fig. 1), it selects the scheduling intervals where to perform the mmWave transmissions for each group of neighbors addressed in the same beam (**line 12** of Fig. 1). The mmWave transmitter could follow different strategies to select the scheduling intervals where to address each group of neighboring vehicles. For example, it could prioritize the transmissions to some neighbors based on the (predicted) channel state. The study of these strategies is out of the scope of this paper. This paper considers instead a general approach by which the $B$ mmWave transmissions are scheduled in the $F$ available scheduling intervals following a clockwise order.

## III. EVALUATION SCENARIO

The performance of the proposed beamwidth-aware scheduling scheme is evaluated through simulations using ns-3.26 and leveraging the mmWave implementation available in [8]. The evaluation considers a highway scenario with 4 lanes and a vehicular density of {75, 150} veh./km. In the scenario, a ratio of {10, 50}% of the vehicles are randomly chosen as mmWave transmitters. We consider that these vehicles want to communicate with all their neighboring vehicles located under LOS conditions and at a distance shorter than 50m. We will refer to them as neighboring vehicles. Without loss of generality, this study considers that IEEE 802.11p (at 6Mbps and 15dBm transmit power) is used as the sub-6GHz V2X technology. All vehicles in the scenario generate sub-6GHz V2X broadcast awareness messages every 100ms. The mmWave scheduling period is then set to $I$=100ms. This study considers that the scheduling period is divided into 5 mmWave scheduling intervals of $T_s$=20ms each. MmWave transmitters schedule new mmWave transmissions to their neighboring vehicles every scheduling period. Following the traffic pattern of [5] for 'collective perception of environment', a mmWave transmitter sends 250 packets of 1600bytes each to the scheduled neighboring vehicle(s) during each scheduling interval. To this aim, the mmWave transmitters use an IEEE 802.11ad-based physical layer at a data rate of 693Mbps and transmission power of 10dBm. The mmWave antenna is initially configured with 60 sectors which can be grouped to form antennas with 30, 20, 15, 12, 10, 6, 5, 4, 3, 2 and 1 sectors. This corresponds to beamwidths of 6º, 12º, 18º, 24º, 30º, 36º, 60º, 72º, 90º, 120º, 180º and 360º, respectively. mmWave receivers use always 6º beamwidth.

## IV. Results

The performance of the proposed beamwidth-aware scheduling scheme is here compared against a baseline scheme in which the mmWave transmitters cannot schedule multiple neighbors in the same scheduling interval, and they use a fixed 6º beamwidth for their transmissions. Like this paper's proposal, the mmWave transmitters that implement the baseline scheme do not use the scheduling intervals where they have been addressed to receive data from other mmWave transmitters (**line 3** of Fig. 1).

Fig. 2 represents the ratio of neighboring vehicles that mmWave transmitters contact in the scheduling periods. The obtained results are represented using a box plot where the red line within the box indicates the median, the edges of the box are the 25th and 75th percentiles, and the whiskers extend to the 5th and 95th percentiles. Results are reported for the evaluated scenarios with {75, 150} veh./km and {10, 50}% of mmWave transmitters; indicated in the figures as (75, 10), (75. 50), etc. In the scenario with a vehicular density of 75 veh./km mmWave transmitters have on average 4.9 neighboring vehicles. For the case of the baseline scheme, the mmWave transmitters could then utilize the 5 configured scheduling intervals to contact their neighbors. However, the results reported in Fig. 2 show that this was not actually the case in 95% of the scheduling periods for the scenario with 10% mmWave transmitters (i.e., (75, 10)). Actually, the mmWave transmitters contacted less than 96%, 56% and 37% of their neighbors in 75%, 25% and 5% of the scheduling periods, respectively; the median value of the ratio of contacted neighbors is 80% in this scenario. This is the case because mmWave transmitters with more than 5 neighbors have not enough scheduling intervals to address them. Some mmWave transmitters may have also less than 5 scheduling intervals available when they are addressed to receive data from other mmWave transmitters (see **line 3** of Fig. 1). Fig. 2 also shows that the ratio of contacted neighbors decreases for the baseline scheme with the increasing vehicular density and ratio of mmWave transmitters. When the vehicular density is 150 veh./km, mmWave transmitters have on average 7.9 neighbors. mmWave transmitters implementing the baseline scheme would require on average 8 scheduling intervals to contact their neighbors. In addition, the increasing ratio of mmWave transmitters results in that it is more likely that mmWave transmitters have other mmWave transmitters as neighbors. This reduces the number of available scheduling intervals that a mmWave transmitter can use to contact its neighbors. Then, in the scenario (150, 50) the mmWave transmitters implementing the baseline scheme contact less than 81% and 31% of their neighbors in 95% and 5% of the scheduling periods, respectively; the median value of the ratio of contacted mmWave neighbors is 40% in this scenario.

Fig. 2 also reports the results obtained for the proposed beamwidth-aware scheduling scheme. In this case, the obtained results show that mmWave transmitters implementing the proposal could schedule mmWave data transmissions to all their neighbors for all the considered scenarios. The proposed scheme benefits of its capability of adapting the mmWave antenna's beamwidth to cover different group of neighbors in the available scheduling intervals. Fig.

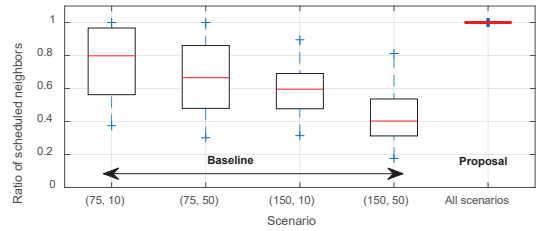

Fig. 2. Box plot of the ratio of contacted mmWave neighbors in each scheduling interval.

3 shows the cumulative distribution function (CDF) of the mmWave antennas' beamwidth that mmWave transmitters used in order to contact all neighboring vehicles. For the scenario (75, 10), the mmWave transmitters used for more than 30% of the scheduled transmissions the initial configuration with a beamwidth of 6º. This means that the mmWave transmitters had to adapt the beamwidth for around 70% of the scheduled mmWave transmissions. For example, more than 20% of the scheduled transmissions utilized beamwidths above 60º. Doing this adaptation in the beamwidth for this scenario, mmWave transmitters contacted on average 1.7 neighboring vehicles in each of the available scheduling intervals.

Fig. 3 also reports the obtained results for the scenarios with an increased vehicular density and ratio of mmWave transmitters. As it was shown for the baseline scheme in Fig. 2, these scenarios challenge the mmWave transmitters to schedule transmissions to all their neighboring vehicles. This is the case because mmWave transmitters have a higher number of neighboring vehicles and fewer mmWave scheduling intervals that they can use to schedule their transmissions. Despite these challenges, mmWave transmitters implementing the proposed beamwidth-aware scheduling scheme are able to schedule transmissions to all their neighboring vehicles (Fig. 2). As it is shown in Fig. 3, this is possible thanks to the adaptive beamwidth performed by the mmWave transmitters. For example, in the scenario (150, 10), the mmWave transmitters only used the initial beamwidth of 6º in 7.8% of the scheduled transmissions. In this scenario, more than 40% of the scheduled transmissions utilized beamwidths above 60º. These beamwidth adjustments resulted in that the mmWave transmitters addressed on average 2.3 neighboring vehicles in each of the available scheduling intervals in this scenario. When the ratio of mmWave transmitters also increases, i.e., scenario (150, 50), the available scheduling intervals decrease. This require that the mmWave transmitters increase the beamwidth they use in the available mmWave scheduling intervals to contact all neighboring vehicles. This can be appreciated in Fig. 3 for the scenario (150, 50) where the obtained results show that the mmWave transmitters only used the initial beamwidth of 6º in 2.5% of the scheduled transmissions, and more than 80% of the scheduled transmissions utilized beamwidths above 60º. It is also important to note that for this scenario the mmWave transmitters utilized a beamwidth of 360º or omnidirectional transmission in ~30% of the scheduled transmissions. The beamwidth adjustments for the scenario (150, 50) resulted in that the mmWave transmitters contacted on average 4.5 neighboring vehicles in the available scheduling intervals.

Finally, Fig. 4 shows the measured packet delivery ratio (PDR) using a box plot representation that indicates its mean value and the 5th, 25th, 75th and 95th percentiles. It is important to note that the PDR is only measured for the neighboring vehicles that are contacted by the mmWave transmitter (see Fig. 2). For the baseline scheme, the obtained results show that the mean value of the PDR is 100% for the scenarios where the ratio of mmWave transmitters is 10%. In these scenarios, the measured PDR is below 95% and 91% in less than 5% of the mmWave transmissions (i.e., $5^{th}$ percentile) when the density of vehicles is 75 veh./km and 150 veh./km, respectively. When the ratio of mmWave transmitters grows, the number of simultaneous transmissions between different pairs of vehicles increases. This results in higher interference levels and an increased likelihood of having scheduling conflicts (i.e., that two mmWave transmitters schedule a transmission to the same neighboring vehicle in (part of) a scheduling interval). These effects contribute to the fact that the $5^{th}$ percentile of the PDR decreases to 73% and 62% for the scenarios (75, 50) and (150, 50), respectively. Fig. 4 shows similar trends for the proposed beamwidth-aware scheduling scheme to those reported for the baseline scheme. The proposed scheme achieves mean values of the PDR above 90% in all scenarios; the mean value of the PDR is also 100% when the ratio of mmWave transmitters is 10%. This is the case in spite of the fact the PDR is computed in the proposal for all neighboring vehicles, and only form some neighbors in the baseline scheme (see Fig. 2 where the mean value of the ratio of contacted neighbors for the baseline scheme ranges from 80% to 40% in the considered scenarios). In the scenarios with 50% mmWave transmitters, the $25^{th}$ and $5^{th}$ percentiles of the PDR show that some of the scheduled transmissions by the proposal are more likely to be affected by effects like scheduling conflicts and interference. By adapting the beamwidths, the mmWave transmitters spread the interference over wider areas and reduce the coverage range and antenna gain. However, the benefits of contacting all neighboring vehicles using the proposed scheme can also be appreciated in terms of the aggregated throughput. Table I shows that the proposed beamwidth-aware scheduling scheme significantly increases the aggregated throughput compared with the baseline scheme. The aggregated throughput measures the rate of packets correctly received by the neighboring vehicles in the scenario. The proposal increases on average by approximately 30% and 70% the aggregated throughput compared with the baseline scheme when the vehicular density is 75 and 150 veh./km, respectively.

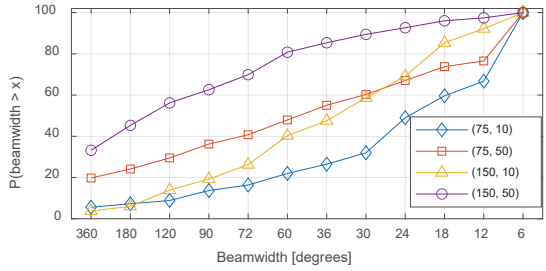

Fig. 3 CDF of the beamwidth utilized in the scheduled mmWave transmissions.

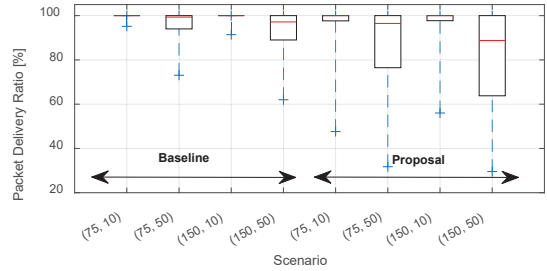

Fig. 4. Box plot of the packet delivery ratio (in [%]) measured over the contacted mmWave neighbors (see Fig. 2).

## V. CONCLUSIONS

This paper has proposed and evaluated a beamwidth-aware scheduling scheme for mmWave V2V communications supported by sub-6GHz V2X technologies. The proposed scheme allows that mmWave transmitters can schedule transmissions to multiple receivers at the same time by adapting the beamwidth. The paper has demonstrated that mmWave transmitters implementing the proposed scheme can schedule transmissions to all their neighboring vehicles in a limited number of scheduling intervals and increase the aggregated throughput compared with a baseline scheme that cannot adjust the beamwidth and address multiple vehicles at the same time. The obtained results have shown the high potential of beamwidth-aware scheduling schemes. To fully exploit the potential of beamwidth-aware scheduling schemes, the effects of beamwidth, antenna gain, coverage area, channel state, scheduling conflicts and interference need to be considered. Future extensions of our proposal could include (all) these effects when mmWave transmissions are scheduled using different weights for each of them depending on the sought objective. To this aim, the support of sub-6GHz V2X technologies can play a critical role.

TABLE I. AVERAGE INCREASE OF THE AGGREGATED THROUGHPUT ACHIEVED WITH THE PROPOSAL COMPARED WITH THE BASELINE SCHEME

| (Vehicular density [veh./km], ratio of mmWave transmitters [%]) | | | |
|---|---|---|---|
| (75, 10) | (75, 50) | (150, 10) | (150, 50) |
| 32.15% | 33.16% | 70.12% | 69.9% |

REFERENCES

[1] P. Kumari, N. J. Myers and R. W. Heath, "Adaptive and Fast Combined Waveform-Beamforming Design for MMWave Automotive Joint Communication-Radar", *IEEE J. Sel. Topics Signal Process., vol. 15, no. 4, pp. 996-1012*, June 2021.

[2] B. Coll-Perales, M. Gruteser and J. Gozalvez, "Evaluation of IEEE 802.11ad for mmWave V2V Communications", *Proc. IEEE WCNC workshop on CmMmW5G*, 15-18 April 2018, Barcelona (Spain).

[3] A. Loch *et al*., "mm-Wave on wheels: Practical 60 GHz vehicular communication without beam training", *Proc. IEEE COMSNETS,* 4-8 June 2017, *Bangalore* (India).

[4] J. Choi *et al*., "Millimeter Wave Vehicular Communications to Support Massive Automotive Sensing", *IEEE Commun. Mag., vol. 54, no. 12, pp. 160-167*, December 2016.

[5] B. Coll-Perales, J. Gozalvez, and M. Gruteser, "Sub-6GHz Assisted MAC for Millimeter Wave Vehicular Communications", *IEEE Commun. Mag., vol. 57, no. 3, pp. 125-131*, March 2019.

[6] A. Molina-Galan, B. Coll-Perales, and J. Gozalvez, "C-V2X Assisted mmWave V2V Scheduling", *Proc. IEEE CAVS 2019*, 22-23 September, 2019, Hawaii (USA).

[7] G. Mendler and G. Heijenk, "On the potential of multicast in millimeter wave vehicular communications", Proc. IEEE VTC2021-Spring, 25-28 April 2021, Helsinki (Finland).

[8] H. Assasa, J. Widmer, "Implementation and Evaluation of a WLAN IEEE 802.11ad Model in ns-3", Proc. ACM WNS3, pp. 57-64, June 2016, Seattle (USA)